\documentclass{article}
\usepackage{spconf,graphicx}
\usepackage{amsmath,amssymb}
\usepackage{url,lipsum,cite}
\usepackage{multicol,multirow}
\usepackage{xpatch,xcolor}
\usepackage{tabularx}
\usepackage{makecell, booktabs}
\newcolumntype{Y}{>{\centering\arraybackslash}X}
\usepackage{hyperref}

\hypersetup{
    colorlinks=true,
    linkcolor=purple,
    filecolor=magenta,      
    urlcolor=purple,
}

\newcommand{\newpara}[1]{\vspace{4pt}\noindent\textbf{#1}}

\title{In search of strong embedding extractors for speaker diarisation}
\name{
  \begin{tabular}{c}
  Jee-weon Jung$^1$, Hee-Soo Heo$^1$, Bong-Jin Lee$^1$, Jaesung Huh$^2$,\\Andrew Brown$^2$, Youngki Kwon$^1$, Shinji Watanabe$^3$, Joon Son Chung$^4$ 
  \end{tabular}
}
\address{
  $^1$Naver Corporation, South Korea\\
  $^2$Visual Geometry Group, Department of Engineering Science, University of Oxford, UK\\
  $^3$Carnegie Mellon University, Pittsburgh, PA, USA\\
  $^4$Korea Advanced Institute of Science and Technology, South Korea
 }
\begin{document}
\ninept
\maketitle

\begin{abstract}
Speaker embedding extractors (EEs), which map input audio to a speaker discriminant latent space, are of paramount importance in speaker diarisation.
However, there are several challenges when adopting EEs for diarisation, from which we tackle two key problems.
First, the evaluation is not straightforward because the features required for better performance differ between speaker verification and diarisation. 
We show that better performance on widely adopted speaker verification evaluation protocols does not lead to better diarisation performance.
Second, embedding extractors have not seen utterances in which multiple speakers exist.
These inputs are inevitably present in speaker diarisation because of overlapped speech and speaker changes; they degrade the performance. 
To mitigate the first problem, we generate speaker verification evaluation protocols that mimic the diarisation scenario better. 
We propose two data augmentation techniques to alleviate the second problem, making embedding extractors aware of overlapped speech or speaker change input. 
One technique generates overlapped speech segments, and the other generates segments where two speakers utter sequentially.
Extensive experimental results using three state-of-the-art speaker embedding extractors demonstrate that both proposed approaches are effective.

\end{abstract}

\section{Introduction}
\label{sec:intro}
Speaker diarisation, which solves the problem of ``\textit{who spoke when}'', is widely used for many applications~\cite{anguera2012speaker,park2022review}. 
It separates a multi-speaker audio input into single-speaker segments and assigns speaker labels.
In the majority of recent works, a speaker diarisation system consists of either a combination of sub-systems such as end-point detection, speaker embedding extraction, and clustering~\cite{landini2022bayesian,shum2013unsupervised,garcia2017speaker,kwon2021adapting,xiao2021microsoft,bredin2020pyannote,park2022multi} or an end-to-end deep neural network~\cite{fujita2019end,bullock2020overlap,horiguchi2020end,maiti2021end,zhang2019fully,medennikov2020target,rybicka2022end} where, in this work, we focus on the former.
When composing a speaker diarisation system based upon sub-systems, the speaker embedding extractor (EE), which maps an utterance to a latent space where speakers can be discriminated, plays the most critical role.

\begin{figure}[t!]
  \centering
  \vspace{-10pt}
  \begin{minipage}[t]{\linewidth}
  \centering
  \centerline{
    \includegraphics[width=0.9\columnwidth,trim={0 0 0 0}]{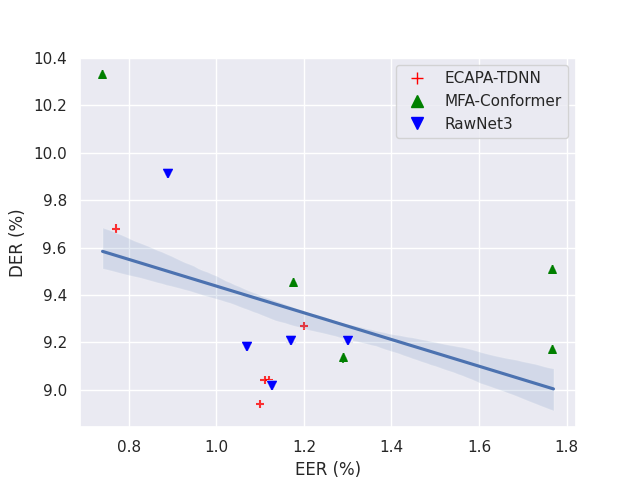}}
  \centerline{(a) Baseline, VoxCeleb1-O}\medskip
\end{minipage}
 \vspace{-25pt}

  \begin{minipage}[t]{\linewidth}
  \centering
  \centerline{
    \includegraphics[width=0.9\columnwidth,trim={0 0 0 0}]{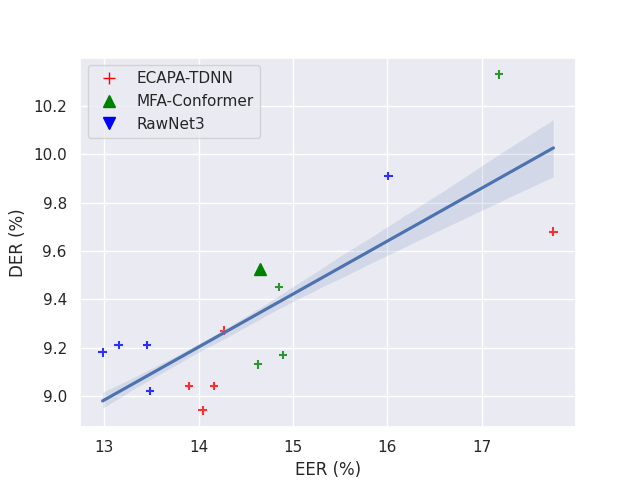}}
  \centerline{(b) Proposed evaluation protocol}\medskip
\end{minipage}

 \vspace{-15pt}
  \caption{
  Correlation between EERs and DERs using three different EEs. 
  Five points for each EE corresponds to five training configurations described in Section~\ref{ssec:exp_res}. 
  DERs are calculated on the VoxConverse test set~\cite{chung2020spot}.
  (a): EERs are calculated on the VoxCeleb1-O test set~\cite{nagrani2017voxceleb}.
  EERs and DERs do not have a positive correlation even though both measures are related to speaker discrimination, and both datasets are from YouTube videos.
  (b): EERs are calculated on the proposed evaluation protocol, described in Section~\ref{sec:eval_protocol}.
  Correlation is higher than (a).
  }
  \vspace{-15pt}
  \label{fig:eer_der}
\end{figure}

In this study, we tackle two problematic phenomena regarding EEs when used for speaker diarisation. 
One is an issue that we raise, and the other is a well-known issue through previous studies~\cite{otterson2007efficient,raj2021multi,landini2022simulated}. 
We first raise the issue that evaluating an EE for speaker diarisation is difficult. 
The straightforward approach would be calculating the diarisation error rate (DER) of a speaker diarisation system using each EE.
However, this is time-consuming and also can be affected by other sub-processes, such as clustering.
Thus, an EE which demonstrates low equal error rates (EER), a metric for speaker verification, on a widely adopted evaluation protocol is typically adopted as an alternative.

As shown in Figure~\ref{fig:eer_der}-(a), we find that lower EER does not guarantee lower DER. 
The correlation between EERs and DERs is not positive, which means that one cannot simply select the EE with the best EER and use it for diarisation.

In our analysis, channel diversity is an essential component which evokes this problematic phenomenon. 
When an EE is used for speaker verification, numerous speakers from diverse channels must be distinguished~\cite{chung2019delving}. 
In contrast, when an EE is used for speaker diarisation, speakers from a single channel need to be distinguished in most cases.
We deduce that negative pairs are hence more challenging when an EE is used for diarisation; between a pair of two different speakers' segments, everything except speaker identity is the same because there is no channel difference within an input for speaker diarisation. 
Therefore, we propose to modify and adapt speaker diarisation datasets for constructing verification evaluation protocols, especially generating pairs within each audio session. 
In addition, we also generate evaluation protocols, including utterances with multiple speakers, to analyse how the EE performs when encountering overlapped or speaker change segments.\footnote{The protocols will be made public before the publication.}
Through experiments, we show that EERs measured using the proposed evaluation protocols have a higher correlation with both DERs and Jaccard error rates (JERs), where JER is another metric for diarisation, which measures average diarisation performance between speakers. 

Meanwhile, we also tackle the problem where speaker embeddings are extracted from segments with multiple speakers.
This problem can occur either by actual overlaps in the input audio or because of the sliding window approach. 
In speaker diarisation, an EE extracts speaker embeddings with a homogeneous shift size. Thus, in speaker change points, two speakers can exist (i.e., speaker change segment). 
The majority of EEs only see single speaker segments when being trained. 
Works such as Kwon et al.~\cite{kwon2021adapting} introduce noise-only segments when training an EE. 
However, as far as we are concerned, EEs are not aware of segments with multiple speakers.
Thus, it may not be surprising that an EE extracts malicious embeddings when encountering such segments. 
In our preliminary experiments, we adopted mix-up~\cite{zhang2018mixup} to account for this issue; however, the results were unsatisfactory. 

We propose data augmentation techniques when training EEs to mitigate this problem. 
Two techniques are proposed: overlapped speech augmentation and speaker change augmentation. 
Both operate in mini-batch-level and on-the-fly schemes. 
We empirically show that both methods are effective, especially when extensively overlapped speech and speaker changes exist.  


\section{Speaker diarisation system}
\label{sec:sd_pipeline}
Our speaker diarisation system comprises four sub-systems: end point detection, speaker embedding extraction, feature enhancement, and clustering.
Since this process pipeline is typical for diarisation, our findings in this study can be valid for similar systems as well. 
Further details regarding our diarisation system can be found in~\cite{kwon2021adapting,kwon2022multi}.

\newpara{End point detection.} 
Our end point detection system first extracts 40-dimensional log mel-spectrograms from the input audio with a window size of 25ms and a shift size of 10ms. 
After applying mean normalisation to log mel-spectrograms, it is fed into a convolutional recurrent neural network with a similar architecture with \cite{cao2019polyphonic}. 
A fully-connected layer then projects the outputs into scalars, which are binarised and then used as the results.

\newpara{Speaker embedding extraction.}
EEs extract speaker embeddings from voice regions detected by the end point detector. 
We adopt a sliding window approach, in line with the majority of recent works~\cite{landini2022bayesian,garcia2017speaker,xiao2021microsoft,bredin2020pyannote}, where we set the window size to 1.5s and shift size to 0.5s. 
Various models can serve as EEs in a speaker diarisation pipeline. 
We adopt three recent state-of-the-art models throughout this study to demonstrate that both the problematic phenomenon and our proposed methods are valid across several models: RawNet3~\cite{jung2022pushing}, ECAPA-TDNN~\cite{desplanques2020ecapa}, and MFA-Conformer~\cite{zhang2022mfa}.
RawNet3 represents models which directly digest raw waveforms; it shows the most competitive performance.
ECAPA-TDNN represents convolution and residual-based models, a widely used variant of the Res2Net~\cite{gao2019res2net}.
MFA-Conformer represents self-attention-based models; it adapts the Conformer~\cite{gulati2020conformer} for speaker verification and demonstrated superior performance and generalisation than CNN-dominant models~\cite{jung2023large}.

\newpara{Feature enhancement.}
We apply dimensionality reduction using an auto-encoder and an attention-based embedding aggregation to refine extracted speaker embeddings adequate for speaker diarisation. 
This process accelerates the clustering step's speed and removes noise in the affinity matrix. 

\newpara{Clustering.}
We adopt both agglomerative hierarchical~\cite{day1984efficient} and spectral clustering~\cite{von2007tutorial,ning2006spectral} to assign speaker labels to each extracted embedding.
Although other configurations are identical to \cite{kwon2021adapting}, we selectively utilise both algorithms based upon the duration of input audio, whereas \cite{kwon2021adapting} selectively adopts one algorithm.

\section{Speaker verification evaluation protocols for diarisation}
\label{sec:eval_protocol}
We find that EERs of EEs, evaluated using a widely adopted evaluation protocol on speaker verification, have a small correlation coefficient with DERs.
Figure~\ref{fig:eer_der} illustrates this phenomenon. 
Even though we use VoxCeleb1-O for EER and VoxConverse test set for DER to minimise the domain gap, it can be seen that the correlation is not sufficient.\footnote{Identical phenomenon also occurs for other speaker diarisation test sets.}
We analyse that this phenomenon has occurred by the different evaluation scenarios between speaker verification and diarisation, as mentioned in Section~\ref{sec:intro}.

\begin{table*}[!t]
  \centering
  \small
  \caption{
    Performances of three state-of-the-art models trained with five different configurations. 
    Four datasets are adopted to report the performances.
    VoxCeleb1-O (Vox1-O) is a widely used speaker verification evaluation protocol and the other three are proposed to simulate how speaker embeddings extractors will perform when adopted in a speaker diarization system (\textbf{Base}: reproduction of original papers, \textbf{Base+}: Base + 1.5s training and noise class, \textbf{OVL}: Base+ with overlapped speech augment, \textbf{SC}: Base+ with speaker change augment, \textbf{Both}: Base+ with both augments). 
  }
  \begin{tabularx}{\linewidth}{l|YYYYY|YYYYY|YYYYY}
    \Xhline{1pt}
    & \multicolumn{5}{c|}{\textbf{RawNet3}} & \multicolumn{5}{c|}{\textbf{ECAPA-TDNN}} & \multicolumn{5}{c}{\textbf{MFA-Conformer}}\\
    \cmidrule{2-16}
    & \textbf{Base} & \textbf{Base+} & \textbf{OVL} & \textbf{SC} & \textbf{Both} & \textbf{Base} & \textbf{Base+} & \textbf{OVL} & \textbf{SC} & \textbf{Both} & \textbf{Base} & \textbf{Base+} & \textbf{OVL} & \textbf{SC} & \textbf{Both}\\
    \Xhline{1pt}
    \multicolumn{16}{c}{\textbf{\textit{Performance on conventional verification evaluation protocol (EER, \%)}}}\\
    \hline\hline
    Vox1-O & 0.89 & 1.13& 1.30& 1.17& 1.07& 0.77 & 1.12 & 1.20& 1.10& 1.11& 0.74 & 1.77 & 1.29 & 1.18 & 1.77\\
    \Xhline{1pt}
    \multicolumn{16}{c}{\textbf{\textit{Performance on proposed verification evaluation protocols (EER, \%)}}}\\
    \hline\hline
    AMI & 13.61 & 11.20& 11.58& 11.44& 11.18& 15.55 & 12.27 & 12.43& 12.40& 12.13 & 15.46 & 13.40 & 12.76 & 12.83 & 12.95\\
    DIHARD3 & 20.36 & 17.59& 17.18& 17.25& 18.04& 22.58 & 17.81 & 17.67 & 18.05& 17.52& 21.81 & 19.81 & 18.92 & 18.92 & 18.38\\
    VoxConverse & 16.01 & 13.48& 13.45& 13.15& 12.98& 17.76 & 14.16 & 14.27 & 14.04& 13.90& 17.18 & 14.89 & 14.63 & 14.85 & 14.62\\
    \hline
    Average & 16.66 & 14.09& 15.31& \textbf{13.94}& 14.06& 18.63 & 14.75 & 14.79 & 14.83& \textbf{14.51}& 18.15 & 16.03 & 15.43 & 15.53 & \textbf{15.31}\\
    \Xhline{1pt}
    \multicolumn{16}{c}{\textbf{\textit{Primary diarization performance (DER, \%)}}}\\
    \hline\hline
    AMI & 20.21 & 19.55& 20.17& 20.08& 19.49& 21.04 & 20.39 & 19.45 & 19.69 & 19.31& 22.22 & 20.98 & 20.78 & 20.68 & 21.36\\
    DIHARD3 & 18.46 & 21.39& 20.51& 17.15& 19.11& 18.29 & 17.88 & 16.24 & 20.63& 16.22 & 18.72 & 21.54 & 22.93 & 21.24 & 19.45\\
    VoxConverse & 9.91 & 9.02& 9.21& 9.21& 9.18& 9.68 & 9.04 & 9.27 & 8.94 & 9.04 & 10.33 & 9.17 & 9.13 & 9.45 & 9.51\\
    \hline
    Average & 16.19 & 16.65& 16.63& \textbf{15.48}& 15.92& 16.33 & 15.77 & 14.98 & 16.42 & \textbf{14.85} & 17.09 & 17.23 & 17.61 & 17.12 & \textbf{16.67}\\
    \Xhline{1pt}
    \multicolumn{16}{c}{\textbf{\textit{Additional diarization performance (JER, \%)}}}\\
    \hline\hline
    AMI & 29.20 & 28.29 & 28.22 & 27.96 & 28.09 & 29.44 & 29.43 & 28.14 & 28.46 & 28.34 & 29.76 & 29.64 & 29.11 & 29.51 & 28.98\\
    DIHARD3 & 43.89 & 45.22 & 45.43 & 43.36 & 44.58 & 43.22 & 43.31 & 42.05 & 44.36 & 42.01 & 43.50 & 44.95 & 45.01 & 45.22 & 44.69\\
    VoxConverse & 36.33 & 35.32 & 35.65 & 35.59 & 35.09 & 36.90 & 35.31 & 35.73 & 34.92 & 35.07 & 36.04 & 34.94 & 35.20 & 35.65 & 35.35\\
    \hline
    Average & 36.47 & 36.27 & 36.43 & \textbf{35.63} & 35.92 & 36.52 & 36.01 & 35.30 & 35.91 & \textbf{35.14} & 36.43 & 36.51 & 36.44 & 36.79 & \textbf{36.34}\\
    \Xhline{1pt}
  \end{tabularx}
  \vspace{-10pt}
  \label{tab:main_res}
\end{table*}

We propose to mitigate this phenomenon by generating and adopting speaker verification evaluation protocols for speaker diarisation, especially for the EE model selection.
The generated protocol is designed to have easier positive and harder negative trials by composing pairs within the same audio file.

Proposed evaluation protocols are generated as follows.
We first crop the input audio into short segments using RTTM~\cite{nist2009rich} files where there exist four types: (i) non-speech, (ii) single speaker, (iii) overlapped, and (iv) speaker change.
Overlapped and speaker change segments here only involve two speaker scenario. 
All segments have a 1.5s duration, identical to the EE's window size.
Then, we compose six types of trials using these segments: (a) target and non-target single speaker-single speaker, (b) target and non-target overlap-single speaker, and (c) target and non-target speaker change-single speaker.
Here, target means that both utterances are from the same speaker.
For target overlapped and speaker change trials, the single speaker corresponds to the major speaker who uttered longer.
For non-target overlapped and speaker change trials, single-speaker does not coincide with any speaker. 
Combining six types of trials, we generate five evaluation protocols to observe and analyse how EE will function when used in a speaker diarisation pipeline:
\begin{itemize}
    \item \texttt{single}: target and non-target trials using only single-speaker segments. 
    \item \texttt{overlap-E}: target and non-target overlap-single speaker trials (easy) are included where overlap ratio is 1\% - 49\%.
    \item \texttt{overlap-H}: target and non-target overlap-single speaker trials (hard) are included where overlap ratio is 50\% - 100\%.
    \item \texttt{speaker change}: target and non-target speaker change-single speaker trials.
    \item \texttt{combined}
    : a combination of the above four protocols.
\end{itemize}

\section{Data augmentation}
\label{sec:augmentation}
We propose two data augmentation techniques to account for EEs when fed overlapped speech and multiple speaker segments from speaker change points. 
Both augmentations are applied at the mini-batch-level. 
Let $\boldsymbol{X} \in \mathbb{R}^{N \times L}$ be a mini-batch, where $N$ and $L$ are the size of mini-batch and utterances' sequence length. 
We set each mini-batch to have at most one utterance per speaker. 
Then, $\boldsymbol{X}'$ is generated by shuffling batch indices of $\boldsymbol{X}$. 
When applying augmentations, utterances in $\boldsymbol{X}$ are used as major speakers with longer durations, whereas utterances in $\boldsymbol{X}'$ are used as minor speakers with shorter durations.

\subsection{Overlapped speech augmentation}
\label{ssec:ovl_aug}
Overlapped speech augmentation adds a minor speaker's scaled and cropped utterance on top of a major speaker's utterance. 
First, we generate $\boldsymbol{X}'_\textrm{cropped}$ by masking random region(s) of $\boldsymbol{X'}$ to zero. 
The duration of the unmasked region is also randomly selected between 200ms and 700ms. 
$\boldsymbol{X}'_\textrm{cropped}$ can have an unmasked region either in the start, end, or middle of an utterance.
Then, $\boldsymbol{X}'_\textrm{cropped,scaled}$ is derived by further scaling $\boldsymbol{X}'_\textrm{cropped}$ to a randomly selected target SNR ratio compared with $\boldsymbol{X}$. 
Finally, augmented mini-batch $\hat{\boldsymbol{X}}$ is derived by adding $\boldsymbol{X}'_\textrm{cropped,scaled}$ to $\boldsymbol{X}$.
Formally, overlapped speech augmentation can be described as:
\begin{equation}
\vspace{-5pt}
\hat{\boldsymbol{x}_i}= \boldsymbol{x}_i + \boldsymbol{M} \otimes \boldsymbol{x'}_i,
\end{equation}
where $\hat{\boldsymbol{x}_i}$, $\boldsymbol{x}_i$, and $\boldsymbol{x'}_i$ are $i^{th}$ utterance of $\hat{\boldsymbol{X}}$, $\boldsymbol{X}$, and $\boldsymbol{X}'$. $\boldsymbol{M} \in \mathbb{R}^L$ is a mask that crops and scales where the values are non-zero for selected crop regions and 0 for the others.

\subsection{Speaker change augmentation}
\label{ssec:sc_aug}
Speaker change augmentation replaces a random region of a major speaker's utterance with a scaled and cropped minor speaker's utterance.
We first select the type of speaker change among three types for each mini-batch: (i) major to minor speaker, (ii) minor to major speaker, and (iii) major to minor to major speaker. 
Then, we derive $\boldsymbol{X}'_\textrm{cropped,scaled}$ in the same fashion with overlapped speech augmentation, using less maximum duration of 300ms. 
A lower maximum duration is designed to counteract speaker change augmentation excessively removing major speaker's information. 
Formally, speaker change augmentation can be described as:
\begin{equation}
\vspace{-5pt}
    \hat{\boldsymbol{x}_i} = \boldsymbol{N} \otimes \boldsymbol{x}_i + \boldsymbol{M} \otimes \boldsymbol{x'}_i,
\end{equation}
where $\boldsymbol{N} \in \{0,1\}^L$ is defined as follows:
\begin{equation}
\vspace{-5pt}
    n_j =
    \begin{cases}
      0, & \text{if}\ m_j > 0 \\
      1, & \text{otherwise},
    \end{cases}
\end{equation}
where $m_j$ and $n_j$ are the $j$th element of $\boldsymbol{M}$ and $\boldsymbol{N}$, respectively.

\section{Experiments}
\label{sec:experiments}

\subsection{Training Datasets}
\label{ssec:trn_db}
We adopt the development sets of the VoxCeleb1\&2 datasets~\cite{nagrani2017voxceleb,chung2018voxceleb2} as the training set.
It comprises 1.2 million utterances from 7,205 speakers, which accounts for approximately 2.7k hours of speech. 

\subsection{Evaluation Datasets}
\label{ssec:eval_db}
We measure the performance using the test or evaluation sets of AMI, DIHARD3, and VoxConverse~\cite{carletta2007unleashing,ryant2020third,chung2020spot}. 

\newpara{AMI evaluation set.}
We adopt the official evaluation partition of the AMI Mix-Headset audio files~\cite{carletta2007unleashing}. 
It comprises diverse meeting scenarios.

\newpara{DIHARD3 evaluation set.}
We use the DIHARD3 full evaluation set to report performances.
This set includes data from various domains, including audiobooks, restaurants, and interviews~\cite{ryant2020third}.

\newpara{VoxConverse test set.}
We use the VoxConverse test set v0.0.2~\cite{chung2020spot}.
This dataset is collected from ``in the wild'' YouTube videos, including multi-media domain data.

\subsection{Models}
\label{ssec:model}
We utilise three state-of-the-art models: RawNet3~\cite{jung2022pushing}, ECAPA-TDNN~\cite{desplanques2020ecapa}, and MFA-Conformer~\cite{zhang2022mfa}, to verify the proposed evaluation protocols and data augmentation techniques. 
These models have been selected to cover a wide range of architectures. 
All three models' architectures have been implemented and trained following corresponding recipes from original papers.

We train each model with five configurations corresponding to each model's five columns in Table~\ref{tab:main_res}. 
`\textbf{Base}' is the baseline model trained for speaker verification, reproducing original papers, with no modifications for diarisation. 
`\textbf{Base+}' refers to the model using 1.5s training and noise class on top of `Base'. 
Training with 1.5s matches the window size of diarisation, and learning to discriminate noise class helps EE when countered with non-speech in diarisation.  
`\textbf{OVL}' and `\textbf{SC}' each refer to applying either of the proposed overlapped speech or speaker change augmentations on top of `\textbf{Base+}'.
`\textbf{Both}' applies both proposed augmentations on top of `\textbf{Base+}'.

\subsection{Proposed augmentations}
\label{ssec:cfg}

For the overlapped speech augmentation, we set the target SNR to a range between 0 and 20. 
In the case of speaker change augmentation, the SNR range is between -5 and 15 to include situations where a minor speaker's utterance is louder. 
Proposed data augmentation techniques are applied to half of the mini-batches to let the model also be trained with the original segments. 
When applying both techniques (`\textbf{Both}'), the ratios are 25\%, 25\%, and 50\% for overlapped speech, speaker change, and no augmentation, respectively. 

\begin{table}[!t]
  \centering
  \small
  \caption{
    Effect of two proposed augmentation techniques on different proposed speaker verification evaluation protocols.
  }
  \begin{tabularx}{\linewidth}{l|YYYY}
    \Xhline{1pt}
    & \textbf{Base+} & \textbf{OVL} & \textbf{SC} & \textbf{Both}\\
    \Xhline{1pt}
    \multicolumn{5}{c}{\textbf{\textit{ECAPA-TDNN on AMI (EER, \%)}}}\\
    \hline\hline
    Single & 11.04 & 11.63 & 10.65 & \textbf{10.45}\\
    Overlap-E & 13.61 & \textbf{13.24} & 13.56 & 13.33\\
    Overlap-H & 27.49 & 28.19 & 27.87 & \textbf{26.96}\\
    Speaker change & 10.96 & 11.08 & 11.06 & \textbf{10.78}\\
    \hline
    Combined & 12.27 & 12.43 & 12.40 & \textbf{12.13}\\
    \Xhline{1pt}
    \multicolumn{5}{c}{\textbf{\textit{MFA-Conformer on VoxConverse (EER, \%)}}}\\
    \hline\hline
    Single & 9.60 & 9.27 & 9.53 & \textbf{9.25}\\
    Overlap-E & 12.58 & \textbf{12.13} & 12.83 & 12.33\\
    Overlap-H & 25.01 & 25.25 & 25.36 & \textbf{24.15}\\
    Speaker change & 14.44 & 14.13 & 14.41 & \textbf{14.11}\\
    \hline
    Combined & 14.89 & 14.63 & 14.85 & \textbf{14.62}\\
    \Xhline{1pt}
  \end{tabularx}
  \vspace{-15pt}
  \label{tab:analysis}
\end{table}

\subsection{Results and analysis}
\label{ssec:exp_res}

Table~\ref{tab:main_res} presents the main results.
In all three models, the lowest EER on the widely used VoxCeleb1-O did not lead to the lowest DER.
As shown in Figure~\ref{fig:eer_der}-(a), which plots all 15 columns of Table~\ref{tab:main_res}, the correlation coefficient was negative.
On the other hand, when using the proposed evaluation protocols, we could find the best EE for all three models in terms of average performance on three datasets.
Applying both augmentations showed the best average performance for ECAPA-TDNN and MFA-Conformer; the speaker change augment had the best average performance for RawNet3. 
Figure~\ref{fig:eer_der}-(b) presents the correlation between EERs calculated using the proposed evaluation protocol and corresponding DERs. 
Comparing it with Figure~\ref{fig:eer_der}-(a), it is clearly demonstrated that the proposed evaluation protocols have a higher correlation with actual diarisation performances.
In addition, JERs have the same correlations with EERs on the proposed evaluation protocols, showing that the proposed protocol is valid and robust for both metrics.
We conclude that the proposed evaluation protocols can be an effective measure when selecting EEs for speaker diarisation.

\newpara{Training configurations.}
We observe that matching the window size of a speaker diarisation system and including noise samples in the training phase did not consistently result in DER improvement. 
Among three models, only ECAPA-TDNN's diarisation performance improved comparing `\textbf{Base}' and `\textbf{Base+}' (16.33\% to 15.77\% in average).
When overlapped speech (`\textbf{OVL}') or speaker change (`\textbf{SC}') augmentation was applied alone, improvement was not consistent. 
Regarding average performance on three datasets, overlapped speech augment showed lower DER in RawNet3 and ECAPA-TDNN; speaker change augment improved the performance in RawNet3 and MFA-Conformer. 
However, when both proposed augmentations were applied (`\textbf{Base+}' vs `\textbf{Both}'), average performance increased for all three models.
These results show that the two proposed augmentation techniques are effective across diverse domains, showing the best results when applied together.

\newpara{Detailed analysis.}
In Table~\ref{tab:analysis}, we further present a detailed analysis of when an EE encounters different types of inputs using four additional evaluation protocols. 
Due to the limited space, we show two cases: ECAPA-TDNN evaluated on the AMI test set, and MFA-Conformer evaluated on the VoxConverse test set. 
In both cases, applying both proposed augmentation techniques demonstrated the best results in all three evaluation scenarios except Overlap-E.
For Overlap-E, only applying overlapped speech augmentation showed the best performance, which is understandable.
Once again, improvement was not consistent when only one augmentation was applied.
However, the two techniques were complementary and synergetic when applied together, even for each case. 

\vspace{-3pt}

\section{Conclusion}
\label{sec:conclusion}
We proposed speaker verification evaluation protocols for selecting EEs when used for speaker diarisation.
EERs calculated using the proposed protocols had a higher correlation with actual diarisation performances than EERs of the widely adopted VoxCeleb1-O speaker verification evaluation protocol.
Furthermore, proposed evaluation protocols simulating specific scenarios such as mild or severe overlaps have enabled detailed analysis of how EEs perform in these situations. 
We also proposed two data augmentation techniques to make EEs aware of overlapped speech and speaker change inputs where multiple speakers exist in a segment. 
Through vast experiments, we demonstrated that the two methods are both effective and that they can be complementary.

\clearpage
\bibliographystyle{IEEEbib}
\bibliography{shortstrings,refs}
\end{document}